# Optical absorption of silicon nanowires

T. Xu[1,2], Y. Lambert[2], C. Krzeminski[2], B. Grandidier[2], D. Stiévenard[2*], G. Lévêque[2], A. Akjouj[2], Y. Pennec[2] and B. Djafari-Rouhani[2]

[1]*Key Laboratory of Advanced Display and System Application, Shanghai University, 149 Yanchang Road, Shanghai 200072, People's Republic of China.*

[2]*Institut d'Electronique et de Microélectronique et de Nanotechnologies, IEMN, (CNRS, UMR 8520), Groupe de Physique, Cité scientifique, avenue Poincaré, 59652 Villeneuve d'Ascq, France.*

We report on simulations and measurements of the optical absorption of silicon nanowires (NWs) versus their diameter. We first address the simulation of the optical absorption based on two different theoretical methods : the first one, based on the Green function formalism, is useful to calculate the scattering and absorption properties of a single or a finite set of NWs. The second one, based on the Finite Difference Time Domain (FDTD) method is well-adapted to deal with a periodic set of NWs. In both cases, an increase of the onset energy for the absorption is found with increasing diameter. Such effect is experimentally illustrated, when photoconductivity measurements are performed on single tapered Si nanowires connected between a set of several electrodes. An increase of the nanowire diameter reveals a spectral shift of the photocurrent intensity peak towards lower photon energies, that allows to tune the absorption onset from the ultraviolet radiations to the visible light spectrum.

*Email address: didier.stievenard@isen.fr

I.  INTRODUCTION

Semiconductor nanowires (NWs) have recently attracted a lot of interest in the fabrication of devices based on the photogeneration of free carriers.[1,2,3,4,5] For example, NWs typically have a smaller diameter than thin films made up of equivalent materials and one could expect shorter carrier transit times.[6,7] Such property turns out to be crucial in increasing the speed of photodetectors or the photoconductive gain of solar cells. In addition, it has been demonstrated that arrays of NWs possess appropriate long optical paths for efficient light absorption.[8,9,10,11] Therefore, semiconductor NWs offer the prospect to improve the energy conversion efficiency in optoelectronic devices.

While in a device consisting of a NW array, size fluctuations in the NW length and diameter are inherent, a few experiments have dealt with the size dependence of the photogeneration of carriers. On germanium NWs, it has been shown that light absorption is not just a function of the intrinsic optical materials properties but it also depends on the excitation of leaky-mode resonances (LMRs), engineered by the geometry and the size of the NWs[12]. It has also been demonstrated that the separation efficiency of the charge carrier is related to the extent of the surface space charge layer with respect to the NW diameter in GaN[13] or Ge[14] nanowires. More recently, the light scattering and the optical absorption of Si NWs have been found to vary with the diameter of Si NWs.[15] Such size effects affect the photoconductive gain and stress the need for additional investigations of the NW length and diameter dependence on their optoelectronic properties in order to better control the photosensitivity of future devices.

Here, in a complementary approach, we first address a simulation of the optical absorption of silicon NW versus their diameter based on two different theoretical tools. The results of the simulation are compared with photoconductivity measurements performed on n-type doped Si NWs. Using NWs that are slightly tapered, such experiments allows to verify the diameter dependence of the optical absorption along the growth direction of single wires, while the structural and compositional of the NWs are kept unchanged, as proved by a complementary electrical analysis of the NWs.

II.   THEORETICAL ANALYSIS

II.1 Formalism

Two theoretical tools are used to investigate the absorption of light by Si NW. The first method, based on the Green function formalism, is useful to calculate the scattering and absorption properties of a single or a finite set of NWs. The second simulation tool, based on the Finite Difference Time Domain (FDTD) method,[16,17] is well-adapted to deal with a periodic set of NWs. After a brief presentation of each formalism, we discuss the physical trends of the absorption by NW of various diameters.

We investigated first the single wire problem with the Green's tensor method. This numerical method is well documented in several articles[18,19] and is particularly convenient for the study of localized objects placed in a multilayered 2D or 3D environment. For convenience, we remind that it relies on the resolution of the Lippmann-Schwinger equation:

$$C_{absorption}(z, \omega) = \int_V d\mathbf{r}\, \varepsilon''(\omega)\, E(\mathbf{r}) \qquad (1)$$

where $\omega$ is the pulsation, $k_0 = \omega/c$, $\mathbf{E}$ is the total electric field at point $\mathbf{r}$, $\mathbf{E}_0$ is the incident electric field (electric field at $\mathbf{r}$ in vacuum), and $\mathbf{G(r,r')}$ the Green's function of the multilayered environment between a source point **r'** and the observation point **r**. The integration is restricted to the volume of the nanowire, which is discretized in several polarisable cells. We have computed the absorption inside the nanowire given by:

$$C_{abso}(z,\omega) = \int_V d\mathbf{r}\, \varepsilon''(\omega) E(\mathbf{r}) \qquad (2)$$

where $\varepsilon''(\omega)$ is the imaginary part of the nanowire susceptibility at the pulsation $\omega$.

In the FDTD method, the calculations are performed on a periodic array of uniform Si wires that lay down in a parallel fashion at the interface between two dielectric layers. Each array is defined by the diameter $d$ of the Si NWs and the spacing $a$ between two adjacent wires, as shown in Fig. 1.

The Maxwell's equations are solved by discretizing both time and space and by replacing derivatives with finite differences. Our calculation is performed in a 2D box (along $x$ and $y$ axes) with propagation along the y axis. Perfect Matching Layer conditions are applied at the boundaries $y$ of the box, in order to avoid reflections of outgoing waves.[20] Along the $x$ direction, the unit cell is repeated periodically and the structure is supposed to be infinite along the $z$ direction. Space is discretized in both $x$ and $y$ directions using a mesh interval equal to $\Delta x = \Delta y = 1$nm. The equations of motion are solved with a time integration step $\Delta t = \Delta x/4c$ and a number of time steps equals to $2^{20}$, which is the necessary tested time for a good convergence of the numerical calculation. The incoming pulse is generated at the bottom part of the unit cell, by a current source parallel to the $x$ axis and having a planar profile along the $x$ direction. The current is generated during a short period of time in such a way as to excite the electromagnetic waves in the frequency domain of interest. Then, the reflection and transmission spectra recorded during the time are Fourier-transformed to obtain R and T as a function of frequency and deduce the absorption as A= 1-T-R.

II.2 Theoretical results

In our calculation, the frequency-dependent complex permittivity of silicon is described by the Lorentz oscillator model[21]:

$$\varepsilon(E) = \varepsilon_1(\infty) + \sum_{i=1}^{4} \frac{A_i}{E_i^2 - E^2 + jE\Gamma_i}, \qquad (3)$$

where $\varepsilon_1(\infty)$ is the high frequency dielectric constant, $A_i$, $\Gamma_i$ and $E_i$ are respectively the amplitude, the damping factor and the resonant energy of the $i^{th}$ oscillator. The parameters are taken from reference[22]. They allow fitting the dielectric permittivity of silicon in the range from 0.2 to 4 eV.

Figure 2(a) shows the absorption spectra of 120-nm-diameter silicon wire in air, for both TE (transverse electric, magnetic field parallel to the wire axis) and TM (transverse magnetic, electric field parallel to the wire axis) illuminations. Spectra are dominated by two modes at about 1.9 eV and 2.7 eV, corresponding to the first volume modes of a cylindrical dielectric waveguide, as the dielectric constant is positive in this energy range. The energies of the first TE (red) and TM (black) eigenmodes are indicated by the ticks at the bottom of the figure. Since the wavelengths λ of the photonic modes in the NWs scale proportionally with their diameter d, one can expect that the position of the absorption peaks will move to lower energies when increasing this diameter. We shall discuss below the relation between the threshold for the absorption of light and the diameter *d*. For energy larger than 3.0 eV, the absorption increases independently of the wavelength due to the higher losses related to electronic transitions. The maps of the electric field intensity for the main two peaks in Fig. 2(a) are plotted in Figs. 2(b) and (c). The dissymmetry along the y-axis is due to the incident illumination along the same direction. Indeed, as the quality factor of the mode around is not very large, the structure of the optical modes is difficult to see because of the superposition of the mode with the incident field. Despite this, we can clearly identify the two modes at 1.9 eV as respectively TE01 and TM11 (see ref. 12).

In Fig. 3, we plot the absorption as a function of the photon energy and the diameter $d$ of the wire, averaged on both polarizations of the incident wave. For diameter larger than 20 nm, the successive modes appear, the lower branch (in cyan) corresponding to the TM11 and TE01 modes of the cylindrical waveguide. Let us stress that this picture is qualitatively very similar to the results obtained in ref.12 for the optical absorption of a Ge NW that we can reproduce by our Green function method.

Before going to the case of a periodic array of NWs, we also use the Green function method to calculate the optical absorption of a grating constituted by a finite number N of Si NWs regularly spaced from each other. The results are given in Fig. 4 for NWs of diameter 120 nm and spacing 200 nm. In case of TE plane wave illumination, modes are slightly blue shifted due to the coupling between adjacent NWs, but the spectra remain unchanged when the number of NWs inside the grating is larger than 5: everything happens as if the grating was infinite. For the TM illumination, the situation is more complex, as the lowest energy mode (1.9 eV) splits into several modes for a large number of nanowires. Its average position oscillates, and tends to stabilize for a number of wires greater than 9. We note that the Lorentz-Drude model used for the Si dielectric constant is incorrect for photon energy smaller than 1.2 eV, which explains the small increase in absorption on Fig. 4(b) around this value.

The above results are completed with the calculation of the optical absorption by a periodic array of Si NW under TM illumination using the FDTD method. Figure 5(a) shows the absorption for various diameters of the NWs ranging from 50 to 160 nm with a period $a$=200nm. Overall, the absorption decreases with smaller diameters, which is understandable owing to the smaller volume occupied by the wire. The trends and the position of the absorption peaks are very similar to those obtained above for a single NW. Remarkably, the threshold energy for the absorption of light by the NWs increases with decreasing $d$. For instance, the energy onset starts around 1 eV for a NW diameter of 160 nm and reaches the UV spectral region, showing the first absorption peak at 3.3 eV for wires with a diameter of 50 nm. We believe that this behaviour is a geometrical effect displaying an intimate relation between the wavelength λ of light in silicon at the threshold and the diameter $d$ of the NW. In fact, deduced from our simulations, we find that the ratio $\lambda/d$ versus $d$ is constant as shown in Fig. 5(b). In previous works,[23] a similar trend was observed when the absorption threshold energy was studied as a function of the filling factor. However, the variation of the filling factor can be realized in two ways, either by changing the diameter $d$ of the nanowire or by changing the period $a$.

Finally, in Figure 6 we compare the absorption of light by suspended and supported arrays of NWs (see figure 1). The trends are the same either the incoming medium for light is air or an insulator such as $SiO_2$. However, one can note that the amplitude of the absorption is increased when the NWs are deposited on a $SiO_2$ substrate.

III. EXPERIMENTAL RESULTS

In order to confirm the dependence of the optical absorption onset on the diameter of a nanowire, n-type Si NWs were synthesized on a Si (111) substrate by the vapor-liquid-solid mechanism using the chemical vapor deposition (CVD) technique. Gold droplets were first formed by evaporating a 2 nm thick layer of gold on a silicon surface. The Si surface covered with the gold thin film was then annealed at 700 °C for five minutes in the CVD reactor prior to NWs growth, in order to produce droplets that were subsequently used as catalysts for the growth of the wires. In the final step, the n-type doped Si NWs were grown at a silane partial pressure of 1 mbar, while the total pressure in the chamber was maintained at 10 mbar with $H_2$ and the temperature set at 430°C. The silane flow rate was 300 sccm (standard cubic centimeters per minute), whereas the flow rate of phosphine diluted in $H_2$ was 1 sccm.

Figure 7 (a) gives a typical TEM image of the end of a n-type doped Si NW grown by CVD. To the right of the gold droplet located at the apex of the NW (left part of the TEM image), the shaft appears speckled over a length of 1 nm. The bright features with sizes of a few nanometers correspond to Au-rich clusters. Such transitional region is attributed to the further growth of the NWs from residual silane gas after the interruption of the gas flow in the growth chamber and also the subsequent diffusion of gold due to the high-temperature inertia of the chamber, after silane has been fully pumped away. Then, in the right-hand side region, that corresponds to constant growth conditions, the contrast on the NW shaft is homogeneous and the NW sidewalls are smooth. From the high resolution acquired in the middle of the NW, a high quality of the crystallographic structure is found and the sidewalls do not appear defective. This result is consistent with the properties obtained in our previous work.[24] From atom probe tomography and secondary ion mass spectrometry, it was found that the concentration of phosphorus impurities establishes at $2\times10^{18}$ at.cm$^{-3}$ in the center of the Si NWs and gradually increases towards the surface (typically, $5\times10^{19}$ at.cm$^{-3}$)[24], in agreement with several studies performed on Si NWs grown with similar conditions[25,26]. At the surface of the Si NWs, the doping level may be different since the incorporation rate of impurities on the NW sidewall has been previously found to be different when lateral growth occurs[27].

To perform electrical and photoconductivity measurements, the Si NWs were deposited onto an oxidized silicon surface from an isopropyl alcohol (IPA) suspension. Several contacts to single Si NWs were made by defining electrodes with electron-beam lithography and by subsequently evaporating a 50 nm-thick layer of Ti and 200 nm-thick layer of Au. The electrode having the smallest separation with the seed particle was carefully positionned away from the region where Au clusters reside.

To illustrate the reproducibility of the doping among NWs, dark current-voltages (I-V) curves are given in Figure 8. In Figure 8 (a), I-V curves have been measured for three different Si NWs through two-terminal electrical contacts. All three Si NWs have a mean diameter of 90 nm between both electrodes. Over a few microns length for each NW, the doping level is homogeneous since the measured resistance (i.e. the slope of the I-V curves) are very similar. Figure 8(b) shows a SEM image of a Si NW contacted with four electrodes, two at the top of the wire, two at the bottom. We note that the separation between both pairs of electrodes is comparable. The NW clearly appear tapered in agreement with a slight lateral overgrowth, as reported before.[24] When the the I(V) characteristics between both pairs of electrodes are measured, a slight variation of the resistance is found and vary accordingly to the small variation of distances between both pairs of electrodes. Such a result indicates an homogeneous doping level along the growth axis. Finally, using a back gate electrode, the I-V curves measured between the electrodes placed at the bottom of the wire as a function of the gate voltage confirms the n-type doping of the wire.

Next, photoconductivity measurements were performed using a monochromator source (ORIEL), with a 100 W Xenon light, allowing a spectroscopic analysis from 350 nm to 1100 nm (which means 3.5 to 1.1 eV). The photon flux was calibrated using a thermopile detector with a broad flat spectral response from 200 nm to 50 microns. All the photocurrent curves were therefore normalized. In order to increase the signal/noise ratio, the light flux was chopped at a frequency of 75 Hz and a lock-in detection was used for the photocurrent measurement. Figure 9 shows the photocurrent characteristics measured on a single Si NW. Whatever the wavelength of the illumination is, the characteristics are highly non-linear. First, the photocurrent increases with the bias up to 1.5 volt. For higher voltage values, there is a decrease of the photocurrent up to a voltage of 2 volts. The maximum of the photocurrent intensity is observed at the energy where the slope of the dark current significantly changes, as seen in the inset of Fig. 9. This behaviour is characteristic of a barrier contact (pseudo- Schottky contact with a low barrier)[28] that occurs between the Si NW and one of the electrodes. At low voltages, the photocarriers are accelerated by the local electrical field in the barrier region, whereas at higher voltages, typically equal or greater than the build in potential, the diode turns into flat band conditions. The electric field is lowered and the current is therefore limited by a diffusion mechanism.[14,29,30] Such observation indicates that there is an electric field associated with the contact between the Si NWs and the electrode, as in the case, for example, of a Schottky contact or an n+-n homojunction. Interestingly, the maximum intensity of the photocurrent increases with decreasing wavelength to reach a saturation regime below 375 nm. By normalizing the peak intensity of the photocurrent with the saturation value of the photocurrent, we can compare the optical absorption of the wire, that is related to the photocurrent, as a function of the photon energy, as shown in Fig. 10 for the NW visible in the SEM image. This NW appears slightly tapered with a top part thinner than its base, as it is seen when both insets of Fig. 10 (a) are compared. Figure 10(b) shows the variation of the maximum of the normalized photocurrent measured in the case of this Si NW for the portions of the wire inserted between electrodes A, B, C and D.

## IV. DISCUSSION

In the Figure 10(b), the curves of the maximum of the normalized photocurrent as a function of the wavelength of the incident photons clearly shifts depending on the position of the photoconductivity measurements along the wire. The onset energies for the absorption peak are determined by assuming linear onsets in the normalized photocurrent plots. Therefore, by measuring the photon energy, when the onsets reach a normalized photocurrent of 1, we find a shift of about 300 meV towards a lower energy threshold for NW diameter increasing from 70 nm (portion AB) to 85 nm (portion CD). As our electrical and structural characterizations do not show any strong fluctuations of the doping level along the growth direction or a significant presence of other impurities due to Au diffusion from the seed particle for example, such a result is consistent with the simulation performed above. It yields a spectral shift of 312 meV for the same variation in the diameter of the wire. Finally, measurements on NWs with much smaller diameters are quite difficult to perform for two reasons : first the absorption is very weak, due to a very small volume of silicon; then it is well known that the ionization energy of the dopants increases with decreasing diameters, leading to a transition towards an insulating material[31]. This effect implies less control on the electrical properties of the contact electrodes and much more fluctuations in the photoconductivity measurements.

V. CONCLUSION

In conclusion, we have studied size-effect in the optical absorption of Si NWs. Calculations of the absorption onset for an array of uniform Si NWs shows that the diameter modifies the light energy absorption behaviour. Such behaviour is confirmed by experimental photoconductivity measurements on single Si NWs, where the modulation of the diameter along the growth axis allows to see spectral shifts in the normalized intensity of the photocurrent as a function of the diameter.

Email address: **didier.stievenard@isen.fr**

**Acknowledgements**

This work was supported by the Délégation Générale de l'Arrmement (DGA) under contract REI N° 2008.34.0031 and in part by the Ministry of Higher Education and Research, Nord-Pas de Calais Regional Council and FEDER through the "Contrat de Projets Etat Region (CPER) 2007-2013".

# FIGURE CAPTIONS

**Figure 1**: Schematic representation of the periodic Si NWs structure. The Si NW diameter is d. The lattice parameter a is defined as the distance between two nearest neighboring Si NWs. The input source is placed in the glass substrate and the detector in air.

**Figure 2:** (a) Absorption spectra for a silicon NW of diameter d = 120 nm, in case of TM (black) and TE (red) plane wave illumination. Vertical lines indicate the eigenfrequencies of the Si NW surrounded by air: degenerated TE/TM are in magenta, TM only in black and TE only in red; (b) and (c): maps of the electric field intensity for the two main peaks in Figure 2 appearing at 1.9eV (or λ= 650 nm) and 2.7eV (or λ= 462 nm): (b) TM illumination and (c) TE illumination.

**Figure 3**: Absorption spectra as a function of the NW diameter, for a plane wave in normal incidence, averaged on TE and TM polarizations. The absorption is calculated per volume unit.

**Figure 4**: Optical absorption for a finite set N of Si NWs of diameter 120 nm regularly spaced by 200 nm, for TE and TM plane wave illumination. The absorption is calculated per volume unit.

**Figure 5:** (a) Absorption of nanowires with diameter of 50, 70, 80, 100, 120, 140 and 160 nm and period a = 200nm. (b) shows the evolution of the ratio between the threshold wavelength λ over d versus d.

**Figure 6** : Absorption of nanowires with diameter of d=100nm nm and period a = 200nm when the nanowires are suspended in air (dashed line) or supported by a SiO$_2$ substrate (heavy line).

**Figure 7** (a) Typical TEM image of the apex of a n-type doped Si NW grown by CVD. Scale bar: 80 nm. (b) Lattice-resolved TEM image obtained in the middle of the Si NW to show the straight sidewalls and the absence of Au clusters when the Si NW is grown at high silane partial pressure.

**Figure 8** (a) *I-V* curve measured for three different Si NWs through from two-terminal current-voltage characteristics. All three Si NWs have a mean diameter of 90 between both electrodes. (b) SEM image of a Si NW in electrical contact with four electrodes. Scale bar: 500 nm. (c) *I-V* curves measured between the labelled electrodes shown in (b). *I-V* curves measured between C and D electrodes as a function of gate voltage.

**Figure 9**: Photocurrent-voltage curves as a function of the wavelength of the incident photons on a single NW with a diameter of 100 nm. The wavelength varies from 575 to 350 nm. Inset : Dark current-voltage curve measured on the same wire.

**Figure 10**: (a) SEM view of an isolated Si NW that is connected with five electrodes, labeled A, B, C, D, E, from the top towards the base of the wire. The left and right insets, that correspond to the top part and base of the wire, highlight the tapered shape of the wire. The scale bars correspond to 150 nm. (b) Variation of the maximum of the photocurrent versus the photon energy measured on the portion of the wire inserted between the electrodes.

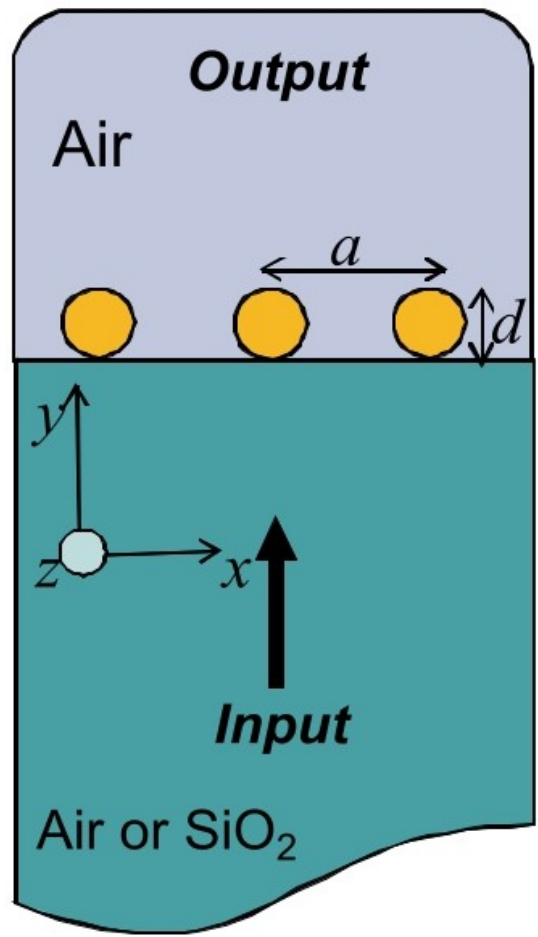

**Figure 1:** Schematic representation of the periodic Si NWs structure. The Si NW diameter is d. The lattice parameter a is defined as the distance between two nearest neighboring Si NWs. The input source is placed in the glass substrate and the detector in air.

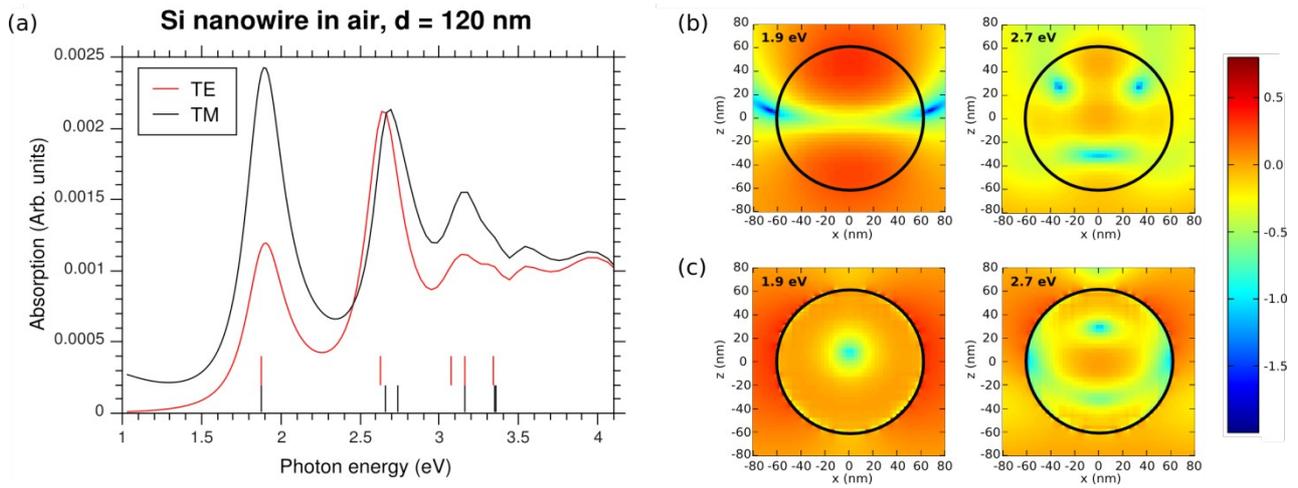

**Figure 2:** (a) Absorption spectra for a silicon NW of diameter d = 120 nm, in case of TM (black) and TE (red) plane wave illumination. Vertical lines indicate the eigenfrequencies of the Si NW surrounded by air: degenerated TE/TM are in magenta, TM only in black and TE only in red; (b) and (c): maps of the electric field intensity for the two main peaks in Figure 2 appearing at 1.9eV (or λ= 650 nm) and 2.7eV (or λ= 462 nm): (b) TM illumination and (c) TE illumination.

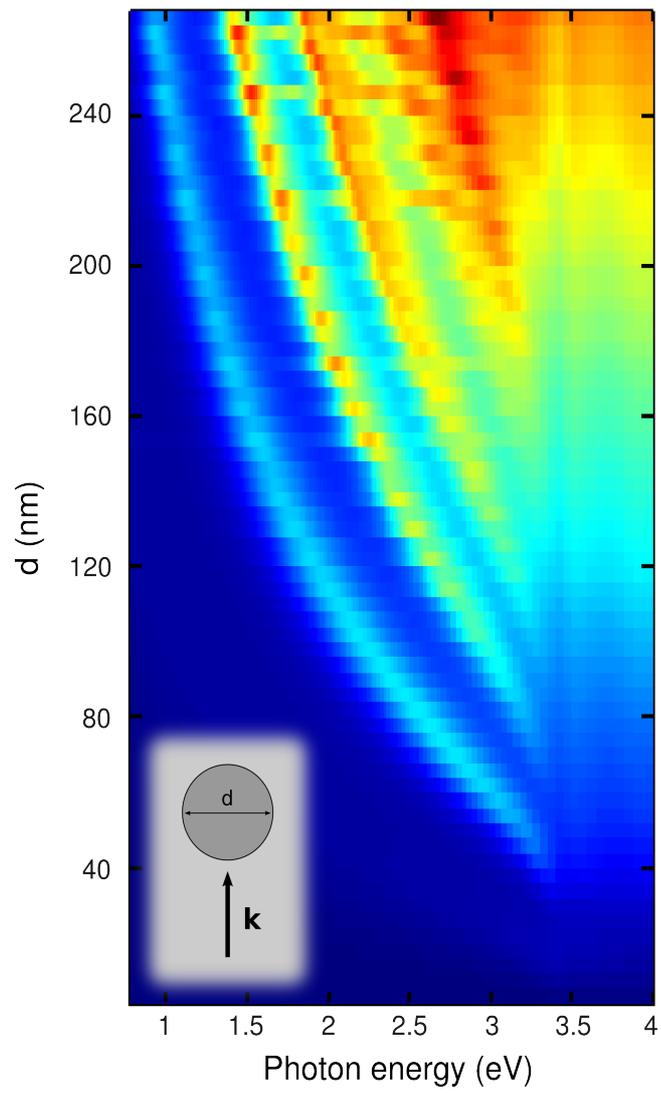

**Figure 3:** Absorption spectra as a function of the NW diameter, for a plane wave in normal incidence, averaged on TE and TM polarizations. The absorption is calculated per volume unit.

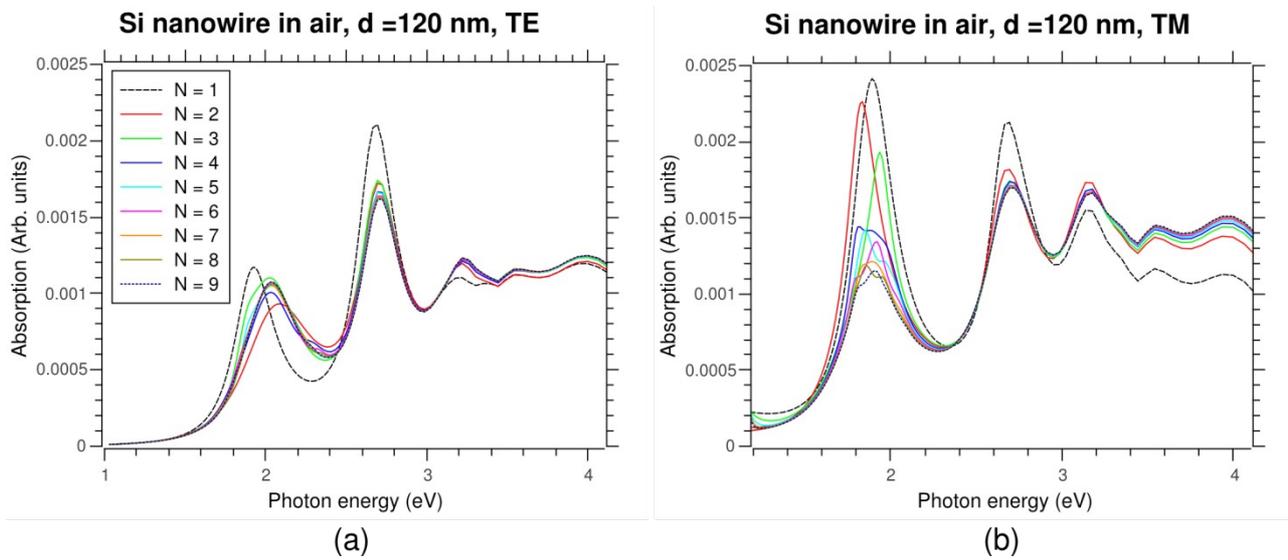

**Figure 4:** Optical absorption for a finite set N of Si NWs of diameter 120 nm regularly spaced by 200 nm, for TE and TM plane wave illumination. The absorption is calculated per volume unit.

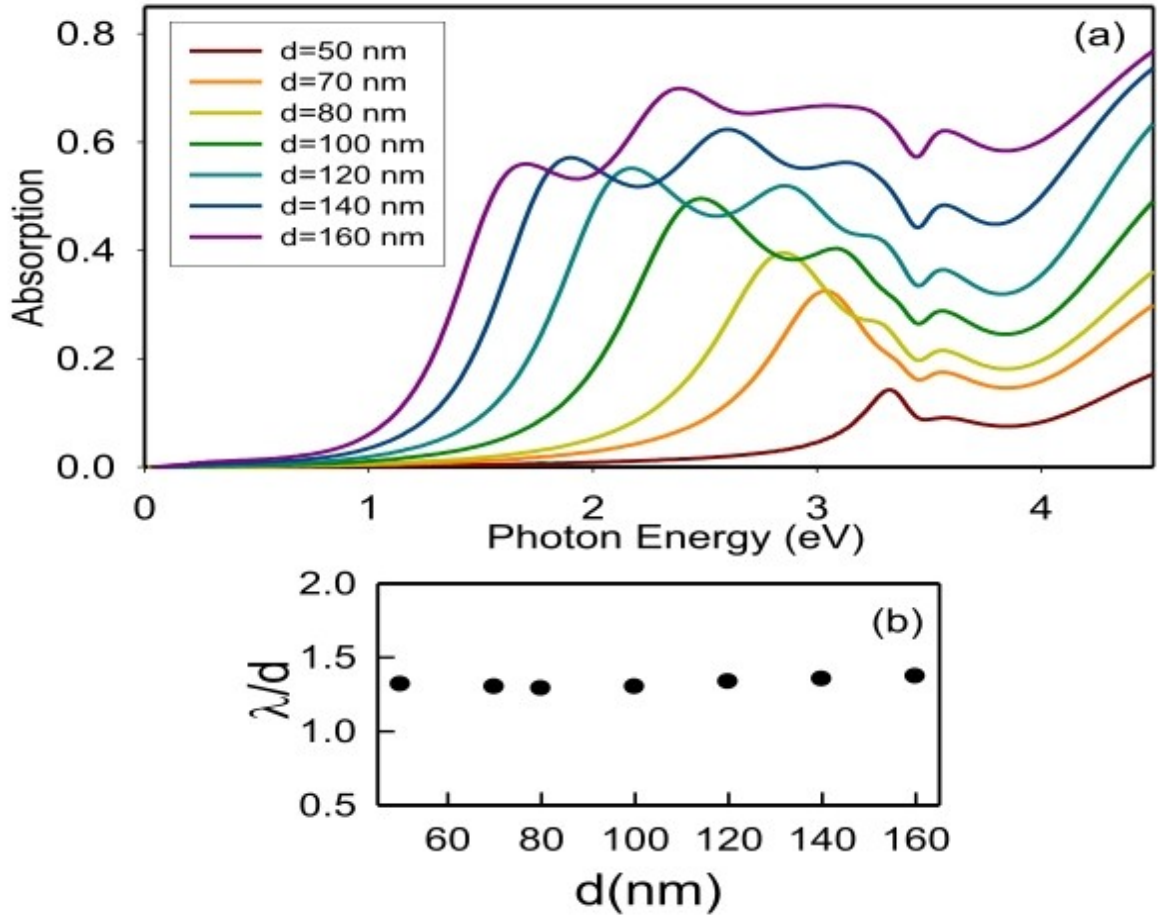

**Figure 5:** (a) Absorption of nanowires with diameter of 50, 70, 80, 100, 120, 140 and 160 nm and period a = 200nm. (b) shows the evolution of the ratio between the threshold wavelength λ over d versus d.

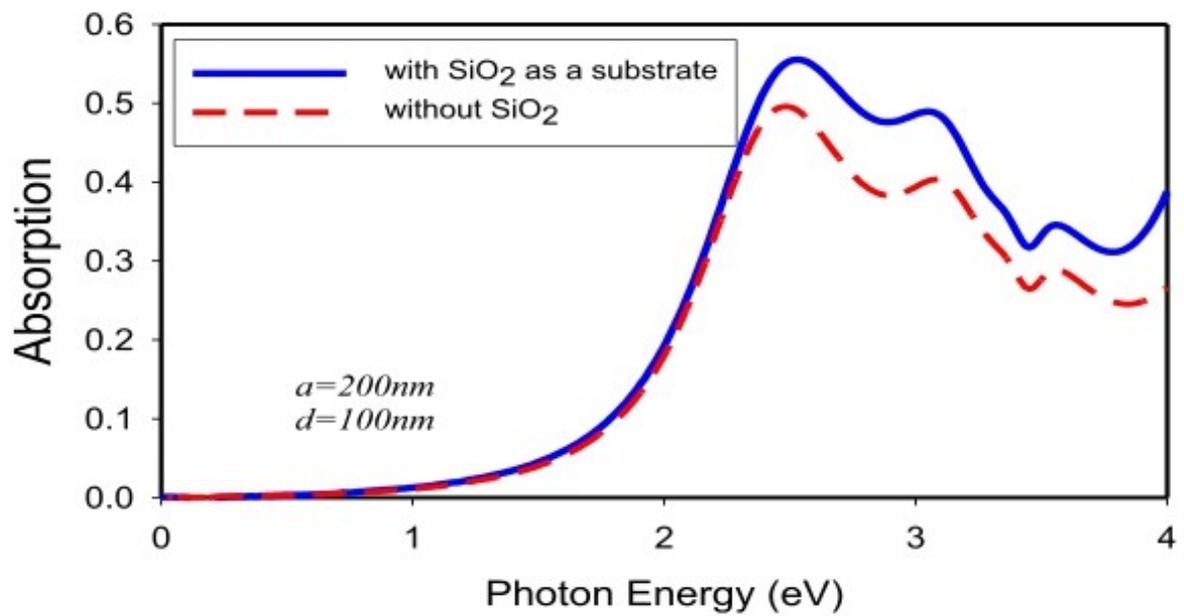

**Figure 6:** Absorption of nanowires with diameter of d=100nm nm and period a = 200nm when the nanowires are suspended in air (dashed line) or supported by a SiO$_2$ substrate (heavy line).

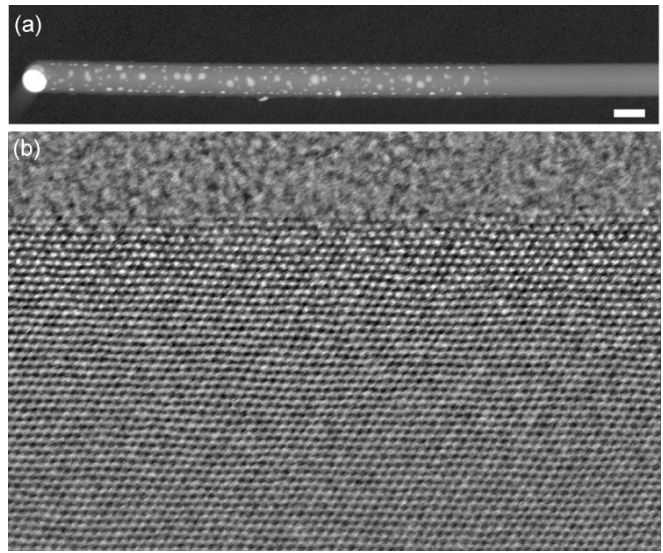

**Figure 7:** (a) Typical TEM image of the apex of a n-type doped Si NW grown by CVD. Scale bar: 80 nm. (b) Lattice-resolved TEM image obtained in the middle of the Si NW to show the straight sidewalls and the absence of Au clusters when the Si NW is grown at high silane partial pressure.

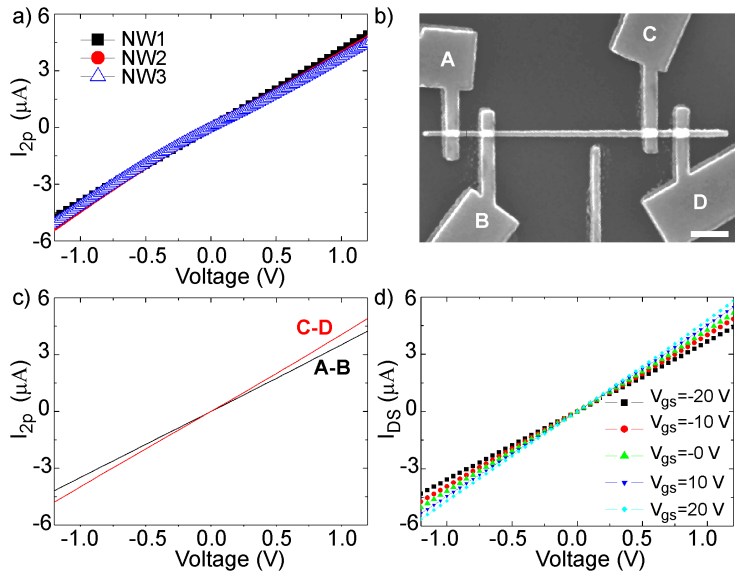

**Figure 8:** (a) *I-V* curve measured for three different Si NWs through from two-terminal current-voltage characteristics. All three Si NWs have a mean diameter of 90 between both electrodes. (b) SEM image of a Si NW in electrical contact with four electrodes. Scale bar: 500 nm. (c) *I-V* curves measured between the labelled electrodes shown in (b). *I-V* curves measured between C and D electrodes as a function of gate voltage.

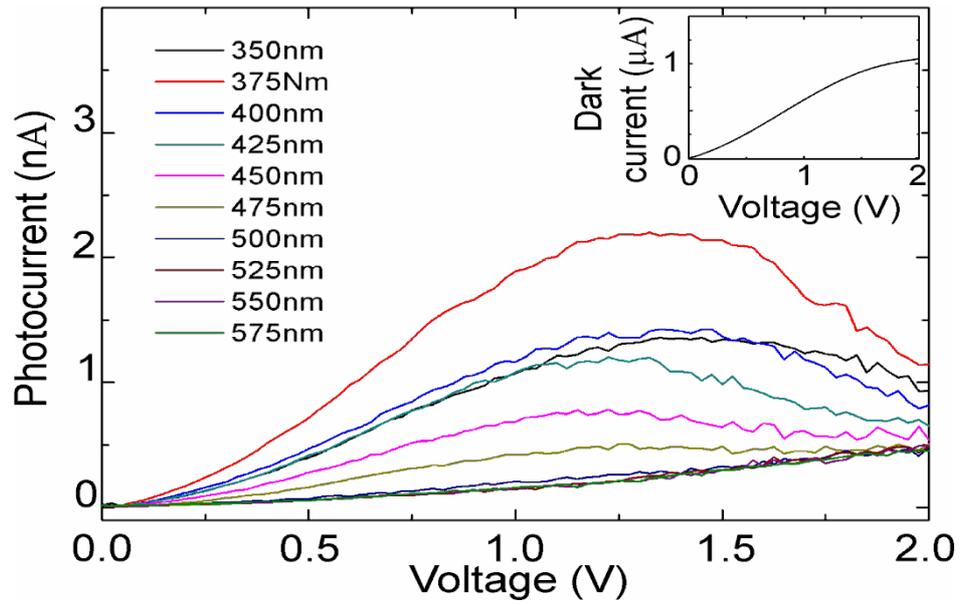

**Figure 9:** Photocurrent-voltage curves as a function of the wavelength of the incident photons on a single NW with a diameter of 100 nm. The wavelength varies from 575 to 350 nm. Inset : Dark current-voltage curve measured on the same wire.

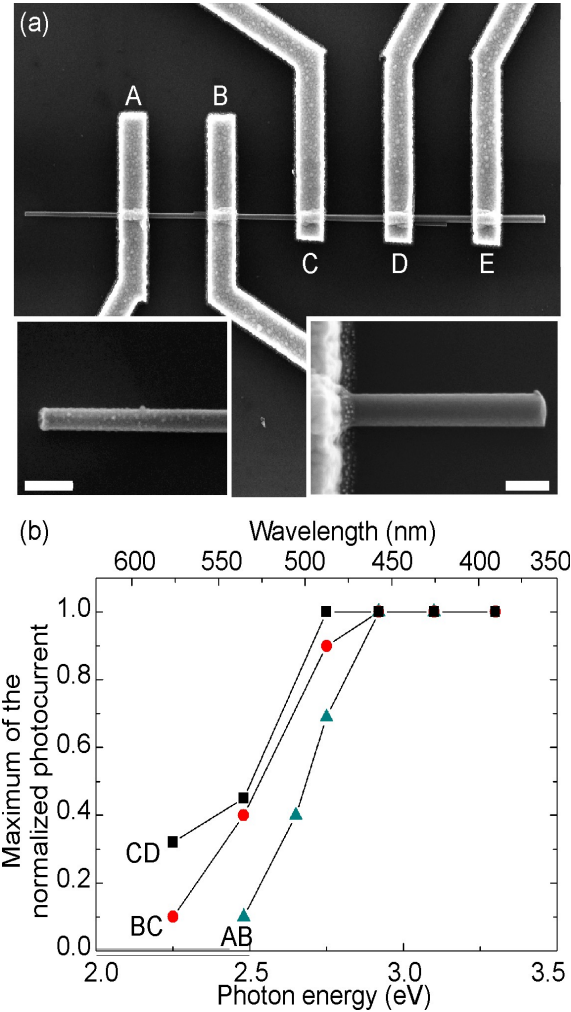

**Figure 10:** (a) SEM view of an isolated Si NW that is connected with five electrodes, labeled A, B, C, D, E, from the top towards the base of the wire. The left and right insets, that correspond to the top part and base of the wire, highlight the tapered shape of the wire. The scale bars correspond to 150 nm. (b) Variation of the maximum of the photocurrent versus the photon energy measured on the portion of the wire inserted between the electrodes.